\documentclass[twocolumn,prl,aps,amsmath,amssymb,showpacs,lengthcheck,floatfix]{revtex4}
\usepackage{graphicx}
\usepackage{epsfig}

\newcommand{\ket}[1]{\vert #1 \rangle}

\newcommand{\ketbra}[1]{\vert #1 \rangle\langle #1 \vert}
\begin{document}
\title{Random bipartite entanglement from W and W-like states}
\author{Ben Fortescue}
\author{Hoi-Kwong Lo}
\affiliation{%
Center for Quantum Information and Quantum Control,
Dept. of Electrical \& Computer Engineering and Dept. of Physics,
University of Toronto, Toronto, Ontario, M5S 3G4, Canada
}%
\date{\today}
\begin{abstract}
We describe a protocol for distilling maximally entangled bipartite states between {\it random} pairs of parties
from those sharing a tripartite $W$ state $\ket{W}=\frac{1}{\sqrt{3}}(\ket{100}+\ket{010}+\ket{001})_{ABC}$, and show that, rather surprisingly,
the total distillation
rate (the total number of EPR pairs distilled per $W$, irrespective of who shares them) may be done at a higher rate
 than distillation of bipartite entanglement between {\it specified} pairs
of parties.  Specifically, the optimal distillation rate for specified entanglement for the $W$ has been previously shown
to be the asymptotic entanglement of assistance of 0.92 EPR pairs per $W$, while our protocol can asymptotically
distill 1 EPR pair per $W$ between random
pairs of parties, which we conjecture to be optimal.  We thus demonstrate a tradeoff between the overall asymptotic
rate of EPR distillation and the distribution of final EPR pairs between parties.
We further show that by increasing the number of parties in the protocol that there exist states with fixed lower-bounded distillable
entanglement for random parties but {\it arbitrarily} small distillable entanglement for specified parties.
\end{abstract}
\maketitle
For pure entangled states $\rho_{AB}$ shared between two parties, Alice and Bob, the standard measure of entanglement is the
Von Neumann entropy $S$
\begin{equation}
S(\rho_{A})=-\rm{tr}(\rho_A\log_2\rho_{A})
\end{equation}
where $\rho_{A}=\rm{tr}_{B}(\rho_{AB})$.
This has been shown to be a fungible measure \cite{BBPS} such that if Alice and Bob occupy
distant laboratories they may, through only local operations in their own laboratories and classical communication 
between their laboratories (LOCC), reversibly convert $N$ copies of $\rho_{AB}$ 
to $NS(\rho_A)$ Einstein-Podolsky-Rosen (EPR) pairs 
\begin{equation}
\ket{EPR}=\frac{1}{\sqrt{2}}(\ket{10}+\ket{01})\label{eq-epr}
\end{equation}
in the large $N$ limit.

For states shared between $>2$ parties the situation is more complex, since there is no single
``maximally entangled state'' (MES) fulfilling the role of the EPR pair in the two-party case.  
One can however consider distillation of multiparty states to EPR pairs shared between two
of the parties.
Previous studies on EPR distillation protocols have focused mainly
on the distillation of EPR pairs between two a priori specified parties.
In contrast, in this paper we consider a different problem---the
distillation of EPR pairs between any (a priori unspecified) pairs of parties.

We find the surprising result that, by not a priori specifying which
pairs of parties share EPR pairs, one can achieve a higher distillation rate of
EPR pairs than what is otherwise possible.
Moreover, we will show that such a surprising result does not
occur for GHZ or certain ``GHZ-like'' states, but does for the $W$-state and certain $W$-like
states. Furthermore, we will also show that, for any $M$-partite pure
state, the regularized relative entropy of
entanglement provides an upper bound on the rate of our random
distillation protocol. We hope that our new line of investigation
presented in this paper will shed some light on the subtleties of
multi-partite entanglement.  Previous results on tripartite and $W$ state distillation include \cite{GLP}, \cite{MB} and \cite{cao}.

We consider distillation of an $M$-party pure state $\psi$ through LOCC
\begin{equation}
\ket{\psi}^{\otimes N}_{A_1\ldots A_m}\longrightarrow 
\bigotimes_{ij}\ket{EPR}_{A_iA_j}^{\otimes N_{A_iA_j}}\label{eq-distill}.
\end{equation}

For specified parties $A_I, A_J$, the {\it asymptotic entanglement of assistance}
(that is, the optimal rate of EPR distillation) $E_{A_IA_J}^\infty(\psi)\equiv \sup_{N\to\infty}\frac{N_{A_IA_J}}{N}$ 
was shown in \cite{HOW} (with the three-party case earlier shown by \cite{SVW}) to be
\begin{equation}
E_{A_IA_J}^\infty(\rho)=\min_T \{S(\rho_{A_IT}),S(\rho_{A_J\overline{T}})\}\label{eq-aea}
\end{equation}
where $\rho=\ketbra{\psi}$ and the minimum is over all partitions of the parties into two groups $T$ and $\overline{T}$.  We
further define
the {\it specified entanglement} $E_s^\infty$ as the maximum of $E_{A_IA_J}^\infty$ over all pairs of parties $I,J$.

We also define the {\it total EPR distillation rate} (the maximum overall rate of distilling
EPR pairs, irrespective of which parties share them) $E_t^\infty(\psi)$ as
\begin{equation}
E_t^\infty(\psi)=\sup\frac{\sum_{ij}N_{A_iA_j}}{N}
\end{equation}
in the limit $N\to\infty$ (thus $E_t^\infty\ge E_s^\infty$ in general).  We further define $E_t$ and $E_s$ as the single-copy
analogues of $E_t^\infty$ and $E_s^\infty$.

We first discuss the case of distilling the $W$ state.
Consider many copies of the $W$ state shared between three parties Alice, Bob and Charlie.
If, say, Bob and Charlie wish to distill EPRs from the $W$s with the help of Alice, then
from (\ref{eq-aea}) we have that the maximum rate (i.e. the maximum number of EPRs per W) which
they can obtain is
\begin{equation}
E_s^\infty(W)=H_2(1/3)\approx 0.92\label{eq-Wentropy}
\end{equation}
where $H_2$ is the binary entropy function
\begin{equation}
H_2(x)=-x\log_2(x)-(1-x)\log_2(1-x).
\end{equation}
By symmetry this is likewise the optimum rate for Alice and Bob distilling EPRs with
Charlie's help etc.  In the case of a single copy of the $W$ state we find from
the general bound of \cite{Gour} that the maximum probability of obtaining an EPR between Alice and Bob
parties is $E_s(W)=G_{AB}(W)=2/3$, where $G_{AB}$ is the concurrence of assistance, originally defined in \cite{LVE}.
(This is in contrast to the GHZ state, for which $E_s=1$ - one can always obtain an EPR between specified parties from
a GHZ through LOCC).

However, suppose the three parties merely wish to distill as many EPRs as possible
without regard for which of the parties share them.  In this case we find they can
achieve a single-copy rate $E_t(W)$, where:\\

\noindent{\bf Theorem 1:
\begin{equation}
E_t(W)\ge 1
\end{equation}}

\noindent{\bf Proof:}
If Alice, Bob and Charlie each apply the rotation
\begin{eqnarray}
\ket{1}\longrightarrow\ket{1},\quad\ket{0}\longrightarrow\sqrt{1-\epsilon^2}\ket{0}+\epsilon\ket{2},\label{unitary}
\end{eqnarray}
then
\begin{multline}
\ket{W}_{ABC}\longrightarrow (1-\epsilon^2)\ket{W}\\
+\frac{\epsilon}{\sqrt{3}}\big(\ket{021}+\ket{201}+\ket{012}+\ket{210}+\ket{102}+\ket{120}\big)\\
+O(\epsilon^2)
\end{multline}

\noindent If all 3
parties then make a measurement on their qubit using the projectors
\begin{equation}
A=\ketbra{0}+\ketbra{1},\quad B=\ketbra{2}\label{povm}
\end{equation}
then either:
\begin{enumerate}
\item All 3 parties get outcome ``A'', with probability $(1-\epsilon^2)^2$, and hence share a $W$ again,
the rotations and projective measurements are then repeated.
\item One of the three parties gets outcome ``B'' (i.e. their qubit is in state $\ket{2}$), with probability $(2/3)\epsilon^2(1-\epsilon^2)$ each.  Say this is Alice,
then following the measurement the state is $\ket{2}_A\otimes\frac{1}{\sqrt{2}}(\ket{01}+\ket{10})_{BC}$
i.e. Bob and Charlie share an EPR pair. 
By symmetry, if the party with a $\ket{2}$ is Bob, then Alice and Charlie will share an EPR pair
and so on, for a total success probability of $2\epsilon^2(1-\epsilon^2)$
\item Two or more parties get outcome ``B'', resulting in a product state, with total probability $\epsilon^4$.
\end{enumerate}
Thus if the parties are performing up to $D$ rounds of the protocol (only performing fewer if an EPR or product state results in fewer than $D$ rounds)
their final expected entanglement is:
\begin{equation}
\langle E_D\rangle = 2\epsilon_D^2(1-\epsilon_D^2)+(1-\epsilon_D^2)^2\langle E_{D-1}\rangle
\end{equation}
where $\epsilon_D$ is the chosen $\epsilon$ for the round of the protocol when up to $D$ rounds remain (thus $\epsilon$
is different in each round).
It follows by differentiation and induction that the optimal $\epsilon_D$ is $\epsilon_D^{opt}=1/\sqrt{D+1}$ which gives
\begin{equation}
\langle E_D^{opt}\rangle=\frac{D}{D+1}.
\end{equation}

Thus for finite $D$ the single copy limit of $E_s=2/3$ is surpassed for $D\ge 3$ and the asymptotic limit
of $E_s^\infty=H_2(1/3)$ is surpassed for $D\ge 12$. In the limit as $D\to\infty$ two of the three parties end up sharing an EPR pair with probability
$\to 1$.  I.e. $E_t^\infty\ge E_t\ge1$. $\Box$\\
This protocol was developed in collaboration with Gottesman \cite{Gott}.

By symmetry, in the limit of many copies $N$ of the $W$ state each pair of parties (Alice-Bob, Bob-Charlie, Alice-Charlie)
will end up sharing on average $N/3$ EPR pairs under this protocol.  We note that the parties could then use the EPRs to share through
quantum teleportation \cite{teleport} $N/2$ copies of the GHZ or any other three-qubit state, for an overall distillation rate of 0.5.
However for GHZ states at least this is not optimal - a rate of 0.64 is demonstrated
(and also shown to be optimal under a specified class of protocols) in \cite{SVW}.

We also find that similar distillation can be advantageous for asymmetric $W$-like states:\\

\noindent{\bf Theorem 2:}
Defining a W-like state
\begin{equation}
\ket{W'}=a\ket{100}+b\ket{010}+c\ket{001}:
\end{equation}
For a $W'$ where (without loss of generality) $0\le a\le b\le c$ with $a,b,c$ real:
\begin{equation}
E_t^\infty(W')\ge 1-(1-(a/c)^2)(b^2+c^2)\left(1-H_2\left(\frac{b^2}{b^2+c^2}\right)\right).\label{eq-wbclower}
\end{equation}
It follows for example that $E_t^\infty(W')\ge 1$ for $b=c$.\\
\noindent{\bf Proof:}
The above rate can be achieved by the combination of a filtering protocol and the random $W$ distillation protocol.

If Alice applies the unitary
\begin{equation}
\ket{0}\longrightarrow \frac{a}{c}\ket{0} +\sqrt{1-(a/c)^2}\ket{2},\quad \ket{1}\longrightarrow \ket{1}
\end{equation}
then
\begin{align}
\ket{W'}\longrightarrow& (a\ket{100}+ab/c\ket{010}+a\ket{001})_{ABC}\nonumber\\
&+\sqrt{1-(a/c)^2}\ket{2}(b\ket{10}+c\ket{01})_{ABC}\label{eq-filt1}
\end{align}
Alice then measures her qubit using the projection (\ref{povm}), obtaining either a tripartite state
(first term in (\ref{eq-filt1}), after normalization) or an entangled pair of Von Neumann entropy $H_2\left(\frac{b^2}{b^2+c^2}\right)$
shared between Bob and Charlie.  This latter outcome occurs with probability $(1-(a/c)^2)(b^2+c^2)$.

We will now show that, in all other circumstances, an EPR pair
is obtained, thus proving the theorem.  If Alice announces
that a tripartite state has been obtained, Bob applies the unitary
\begin{equation}
\ket{0}\longrightarrow \frac{b}{c}\ket{0} +\sqrt{1-(b/c)^2}\ket{2},\quad \ket{1}\longrightarrow \ket{1},
\end{equation}
thus leaving the three parties with the state
\begin{eqnarray}
\ket{\psi}&=&\frac{1}{\sqrt{2+(b/c)^2}}\Bigg(\frac{\sqrt{3}b}{c}\ket{W}_{ABC}\nonumber\\
&&+\sqrt{2}\ket{2}_B\sqrt{1-(b/c)^2}\ket{EPR}_{AC}\Bigg).
\end{eqnarray}
Bob performs the projection (\ref{povm}) to obtain either a shared $W$ or a shared EPR between Alice and Charlie.
Bob announces his result - if a $W$ is obtained then the random $W$ distillation is performed
to obtain a randomly shared EPR pair. $\Box$

For the $W'$, $E_s(W')=H_2(b^2)$ which is less than or equal to the lower bound on $E_t^\infty$ of Theorem 2.\\

\noindent{\bf Conjecture: $E_t^\infty(W)=1$}\\
That is, we conjecture that distillation using the protocol described in Theorem 1 is optimal
for the $W$ state.
However we have no proof of this - our tightest upper bound is as follows:\\

\noindent{\bf Theorem 3:}
For a pure tripartite state $\sigma_{ABC}$
\begin{align}
E_t^\infty(\sigma_{ABC})\le \min\{S(\sigma_{BC})+E_r^\infty(\sigma_{BC}),\nonumber\\
S(\sigma_{AC})+E_r^\infty(\sigma_{AC}),S(\sigma_{AB})+E_r^\infty(\sigma_{AB})\}\label{eq-limit}
\end{align}
where the asymptotic relative entropy of entanglement $E_r^\infty(\rho)=\lim_{N\to\infty}E_r(\rho^{\otimes N})/N$
and for an $M$-party state
\begin{equation}
E_r(\rho_{A_i\ldots A_M})=\min_{\sigma_{A_i\ldots A_M}^\mathrm{sep}} S(\rho_{A_i \ldots A_M}||\sigma_{A_i\ldots A_M}),
\end{equation}
where $\sigma_{A_i A_j}^\mathrm{sep}$ are separable states\\
\noindent{\bf Proof:}
(Our proof is a simple application of the result in \cite{LPSW}).
It was shown in \cite{LPSW} that for any three-party LOCC protocol starting from
a pure initial state $\rho_{ABC}$
\begin{align}
\langle E_r(\rho_{BC}) \rangle_\mathrm{final}-E_r(\rho_{BC})_\mathrm{initial}\nonumber\\
\le S(\rho_A)_\mathrm{initial}-\langle S(\rho_A)
\rangle_\mathrm{final}\label{eq-er}
\end{align}
For a distillation (\ref{eq-distill}) of a pure state $\sigma_{ABC}$ we have, assuming asymptotic continuity,
\begin{align}
S(\rho_A)_\mathrm{initial}=&S(\sigma_A^{\otimes N})=NS(\sigma_A)\\
\langle S(\rho_A) \rangle_\mathrm{final}=&N_{AB}+N_{AC}\\
\langle E_r(\rho_{BC}) \rangle_\mathrm{final}=&N_{BC}\\
E_r(\rho_{BC})_\mathrm{initial}=&E_r(\sigma_{BC}^{\otimes N})
\end{align}
thus
\begin{align}
N_{AB}+N_{BC}+N_{AC} \le& NS(\sigma_A)+E_r(\sigma_{BC}^{\otimes N})\nonumber\\
=&NS(\sigma_{BC})+E_r(\sigma_{BC}^{\otimes N})
\end{align}
Since we are free to permute $\{A,B,C\}$,
dividing through by $N$ and taking $\lim_{N\to\infty}$ leads to (\ref{eq-limit}). $\Box$

\begin{figure}
\includegraphics[width=6.5cm]{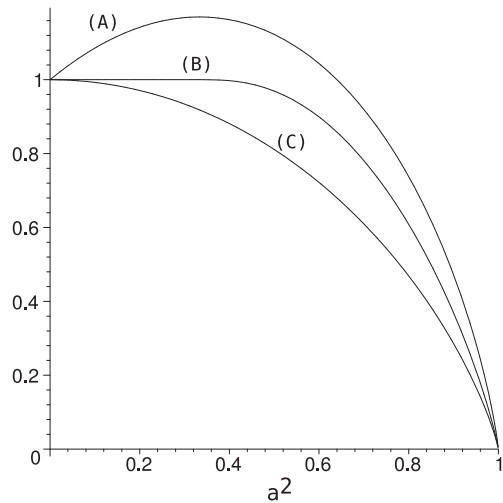}
\caption{For $W_{ab}$, a plot as a function of $a^2$ of
(A) Upper bound on $E_t^\infty$ (as specified in Eq. (\ref{eq-wbcbound})),
(B) Lower bound on $E_t^\infty$ (as specified in Eq. (\ref{eq-wbclower})),
(C) $E_{s}$ (``specified entanglement''), equal to $H_2(b^2)$.
The gap between (B) and (C) shows that distillation to random parties can be more
efficient by certain measures than distillation to specified parties.}\label{fig-wplot}
\end{figure}

Theorem 3 leads to an explicit bound on $E_t^\infty$ for states defined as
$\ket{W_{ab}}\equiv a\ket{100}+b\ket{010}+b\ket{001}$ ($a$, $b$ real).  From \cite{VP} (Eqs (54)-(56)) we have that for $W_{ab}$
\begin{equation}
E_r(\sigma_{BC})_\mathrm{initial}=-(1+a^2)\log_2\left(\frac{1+a^2}{2}\right)+a^2\log_2 a^2.
\end{equation}
Since $E_r(\sigma_{BC})\le E_r^\infty(\sigma_{BC})$ and $S(\sigma_A)=H_2(a^2)$ we have
\begin{align}
E_t^\infty(W_{ab})\le& -(1-a^2)\log_2(1-a^2)\nonumber\\
&-(1+a^2)\log_2\left(\frac{1+a^2}{2}\right).\label{eq-wbcbound}
\end{align}
This is illustrated in Figure \ref{fig-wplot}.  This bound is a maximum for the $W$ state with $a^2=1/3$,
for which $E_t^\infty(W)\le\log_2(9/4)\approx 1.17$.

We also find a more general bound for any number of parties:\\

\noindent{\bf Theorem 4:}
For an $M$-party pure state $\sigma_{A_1\ldots A_M}$.
\begin{equation}
E_t^\infty(\sigma) \le E_r^\infty(\sigma)\label{eq-erbnd}
\end{equation}
(We thank Martin Plenio for pointing out this bound to us in the tripartite case, which follows from Theorem 3 above and Theorem 1 of \cite{PV}).

\noindent{\bf Proof:}
\cite{PV} derives a bound on the relative entropy of tripartite systems from \cite{PVP}, noting that this readily
generalizes to the multiparty case.  The general multiparty bound is
\begin{align}
E_r^\infty(\sigma_{A_1\ldots A_M})\ge& \max\{S(\sigma_{A_1\ldots A_{M-1}})\nonumber\\
&+E_r^\infty(\sigma_{A_1\ldots A_{M-1}}),\ldots\}
\end{align}
where the maximum is over all permutations of the parties $A_1$ to $A_M$.
Considering the final state in (\ref{eq-distill}) $\rho^f_{A_1\ldots A_M}=\bigotimes_{ij}\ket{EPR}_{A_iA_j}^{\otimes N_{A_iA_j}}$, we have:
\begin{equation}
S(\rho^f_{A_1\ldots A_{M-1}})=\sum_i N_{A_i A_M}
\end{equation}
and, by induction from the three-party bound
\begin{equation}
E_r^\infty(\rho^f_{A_1\ldots A_{M-1}})\ge\sum_{\{i,j\}\ne M}N_{A_i A_j}
\end{equation}
Thus (since $E_r^\infty$ is an entanglement monotone), for the distillation (\ref{eq-distill}),
$N\times E_r^\infty(\psi)\ge E_r^\infty(\psi^{\otimes N})\ge E_r^\infty(\rho_f) \ge \sum_{ij} N_{A_i A_j}$, leading to (\ref{eq-erbnd}). $\Box$

Various conclusions follow from this bound - since $E_r^\infty(\rho_{ABC})\le E_r(\rho_{ABC})$ generally, we find for example
that since for GHZ-like states $\ket{GHZ'}=\alpha\ket{000}+\beta\ket{111}$ we have
$E_r^\infty(GHZ')=E_s(GHZ')=H_2(|\alpha|^2)$ \cite{PV}, then random distillation gives no advantage over specified distillation for such states.

The bound also leads to the same numerical bound for $W$ as above, as shown in \cite{GPV} which gives $E_r^\infty(W)\le E_r(W)=\log_2(9/4)$.  In addition, since \cite{IP}
showed that $E_r^\infty(W)\ge\log_2 3-5/9\approx 1.03$, any numerical upper bound on $E_t^\infty(W)$ derived from (\ref{eq-erbnd}) cannot be less than 1.03 and hence
would not be sufficient in itself to prove our conjecture.

Our protocol for $W$ states (in which a randomly determined party announces
their measurement result to leave the remaining two parties with an EPR pair)
can be straightforwardly generalized to a multiparty protocol in which multiple
announcements are made, which leads to the following result:\\

\noindent{\bf Theorem 5:}
One can construct states with arbitrarily small $E_s^\infty$ for which $E_t^\infty\ge1$.

\noindent{\bf Proof:}
Consider the class of states which we denote as $\ket{W_M}$:
\begin{equation} 
\ket{W_M}=\frac{1}{\sqrt{M}}(\ket{00\ldots01}+\textrm{cyclic permutations})
\end{equation}
(so $W_2$ is an EPR pair, $W_3$ is a $W$ etc.)  The $W_M$ state is initially shared between $M$ parties, all of whom perform the unitary (\ref{unitary})
 on their
qubit, followed by the projection (\ref{povm}), repeating as necessary until one party gets outcome ``B'',
as with the $W$.  This party announces their result and the remaining parties repeat the protocol.

After one successful application of the protocol one party has made an announcement and the remainder
share an $W_{M-1}$ state and so on.  After $M-2$ such rounds the 2 remaining parties share an EPR pair, thus
\begin{equation}
E_t(W_M)\ge 1
\end{equation}
but for  a $W_M$ state
\begin{equation}
E_s^\infty(W_M)=H_2(1/M)
\end{equation}
which $\to 0$ as $M\to\infty$. $\Box$

In future, clearly we would like to prove or disprove our conjecture regarding the optimality of the random
distillation for the $W$ state by finding a tight upper bound for $E_t^\infty$, as well as tightly bounding
$E_t^\infty$ for more general tripartite states.  Though our operational measure $E_t^\infty$ is based
on distillation in the many-copy limit, our present random distillation protocols work
on single copies of states - it is not clear whether distillation rates could be improved
by operating on multiple copies.

In addition, a more discriminating quantity for tripartite states is the range of obtainable
values of $\{N_{AB},N_{BC},N_{AC}\}$ in the distillation (\ref{eq-distill})
 - an interesting problem is to tightly bound this range for, say, general $W'$.  It would
likewise be worth investigating the reverse process - the required number of
shared EPRs between parties for formation of $W'$.

So far we have only investigated random distillation of a particular class of pure states.
It would be interesting to study random distillation for other
types of
output states including the $W$ and GHZ states. One might even
study the random distillation and irreversibility in distillation and
formation
between a whole hierarchy of states.
We note that there have been two recent papers on distillation
of mixed stabilizer states (\cite{KMBD},\cite{GKV} - note that the $W$ is not a stabilizer state) - it would be interesting to find the achievable
random distillation rates for such states as well as for more general multipartite states.\\

We thank Matthias Christandl, Andreas Winter
and, particularly, Daniel Gottesman and Martin Plenio for enlightening discussions.
Financial support from NSERC, CIAR, CRC Program, CFI, OIT,
PREA, MITACS and CIPI is gratefully acknowledged.  BF additionally acknowledges support from
a University of Toronto EF Burton Fellowship.

\end{document}